\documentclass[12pt]{article} \textwidth=6in \oddsidemargin=0in
\usepackage{amssymb}
\begin{document}
\def\ov{\over} \def\be{\begin{equation}} \def\ee{\end{equation}}
\def\bc{\begin{center}} \def\ec{\end{center}} \def\noi{\noindent}
\def\s{\sigma} \def\x{\xi} \def\ld{\ldots} \def\cd{\cdots}
\def\({\Big(} \def\){\Big)} \def\f{\mathfrak f} \def\S{\mathcal S}
\def\s{\sigma} \def\t{\tau} \def\ph{\varphi} \def\G{\Gamma} \def\inv{^{-1}}
\def\sp{\vspace{1ex}} \def\wh{\widehat} \newcommand{\xh}[1]{\wh{\x_#1}}
\def\U{\mathfrak U} \def\l{\ell} \def\ni{\not\in} \def\iy{\infty}
\newcommand\tbin[2]{\left[{#1\atop #2}\right]_\t} \def\xm{\x_{m+1}}
\def\sm{\s(m+1)} \def\P{\mathcal P} \newcommand{\xs}[1]{\x_{\s(#1)}}
\newcommand{\ys}[1]{y_{\s(#1)}} \def\ep{\varepsilon} \def\C{\mathcal C}
\def\ds{\displaystyle} \def\bs{\backslash} \def\Z{\mathbb Z} \def\la{\lambda}
\def\PP{\mathbb P}

\hfill July 23, 2017

\bc{\bf\large Blocks in the Asymmetric Simple Exclusion Process}\ec

\begin{center}{\large\bf Craig A.~Tracy}\\
{\it Department of Mathematics \\
University of California\\
Davis, CA 95616, USA}\end{center}

\begin{center}{\large \bf Harold Widom}\\
{\it Department of Mathematics\\
University of California\\
Santa Cruz, CA 95064, USA}\end{center}

\begin{abstract}{In earlier work the authors obtained formulas for the probability in the asymmetric simple exclusion process that the $m$th particle from the left is at site $x$ at time $t$. They were expressed in general as sums of multiple integrals and, for the case of step initial condition, as an integral involving a Fredholm determinant. In the present work these results are generalized to the case where the $m$th particle is the left-most one in a contiguous block of $L$ particles. The earlier work depended in a crucial way on two combinatorial identities, and the present work begins with a generalization of these identities to general $L$.}\end{abstract}

\bc{\bf I. Introduction}\ec

The  \textit{asymmetric simple exclusion process} (ASEP) on the integer lattice $\Z$ is one of the most important stochastic models in nonequilibrium statistical physics and interacting particle systems. For example, in the case of \textit{step initial condition} (particles initially occupying the positive integer sites $\Z^+$), a formula for the
distribution of the $m$th particle from the left  \cite{tw3}
was the starting point for the derivation of a formula for the one-point probability distribution of the height function for the KPZ equation with narrow wedge initial conditions \cite{ACQ,SS}.
\par
Recall that in ASEP a particle waits an exponential time, then moves one step to the right with probability~$p$ if the site is unoccupied (or else stays put) and one step to the left with probability $q=1-p$ if the site is unoccupied (or else stays put). 

The first result in \cite{tw1} was the derivation for $N$-particle ASEP of a  formula for the transition probability $P_Y(X,t)$ from configuration $Y$ to configuration $X$ in time~$t$.   Here  $Y=\{y_1,\ldots, y_N\}\in \Z^N$ with $y_1<y_2\cdots<y_N$ and $X=\{x_1,\ldots, x_N\}\in \Z^N$ with $x_1<x_2<\cdots<x_N$.  The formula for $P_Y(X,t)$ is quite complicated as it is a sum of $N!$ terms
with each term an $N$-dimensional integral. (See (\ref{PXY}) below).  
From the transition probability $P_Y(X,t)$ one can derive expressions for $\PP(x_m(t)=x)$, the probability that at time $t$ the $m$th particle from the left is at site $x$. (Theorems 5.1 and 5.2 of \cite{tw1}.) The derivations depended crucially on identities (1.6) and (1.9) of \cite{tw1}. For step initial condition, with some additional analysis one reduces the one-point function to an integral whose integrand involves a Fredholm determinant. (This is formula (1) of \cite{tw2}).

The stated results were all for the one-point function, and it is natural to try to extend them to joint probabilities,
$\PP(x_{m_1}(t)=x_1,\ld,x_{m_r}(t)=x_r)$.
Some exact formulas for these were derived in \cite{tw4} but they were quite complicated and do not seem to lead to anything useful. Here we consider a special case for which we are able to generalize the main results of \cite{tw1,tw2}.

By a \textit{block of length $L$} (or {\it L-block}, for short) we mean a contiguous block of $L$ particles. We are interested in the probability that at time $t$ the $m$th particle from the left is the beginning of
an $L$-block starting at $x$. Precisely, we want the probability $\P_{L,Y}(x,m,t)$ of the event
\[ x_m(t)=x,\ x_{m+1}(t)=x+1,\ldots,x_{m+L-1}(t)= x+L-1,\]
given the initial configuration $Y$. As mentioned, all the results of \cite{tw1,tw2} and beyond depend on a certain pair of identities, and a stumbling-block to generalizing the results has been finding generalizations of those identities. We have been able to do this for $L$-blocks and we call the generalizations {\it Identity~$1_L$} and {\it Identity~$2_L$}. Once we have them, we are able to follow the paths followed in \cite{tw1,tw2}. Because of this we shall omit some details.

In Section II we derive Identities $1_L$ and $2_L$. In Section III in Theorem 1 we derive a formula for $\P_{L,Y}(x,m,t)$ for $N$-particle ASEP as a sum of multiple integrals over small contours. (The singularities of the integrand are outside the contours.) Theorem~2 gives an analogous expansion where the integrals are over large contours. (The singularities of the integrand are inside the contours.) The large contour expansion extends to  infinite systems with initial conditions unbounded on the right. Finally, in Section IV in Theorem 3 there is an expression for the step initial condition probability $\P_{L,\,\Z^+}(x,m,t)$ as an integral involving a Fredholm determinant.

In future work we hope to obtain asymptotic results as $N\to\iy$ analogous to those in \cite{tw3} for step initial condition. The last-mentioned formula, even in the case $L=1$, is difficult to analyze asymptotically. In \cite{tw3} a deformation theory was developed which made it possible to
replace the kernel in the formula by one with the same Fredholm determinant and which was amendable to asymptotic analysis. (See Theorem 3 in \cite{tw3}). We expect the method to apply to this more general case as well. 

\bc{\bf II. Two identities}\ec

The two identities are expressed in terms of a symmetric polynomial $\f_L(\x_1,\ld,\x_N)$ (or $\f_L(\x)$, for short) defined by (\ref{fLdef}) below. To state them we introduce the notation
\[U(\x,\x')={p+q\x\x'-\x\ov \x'-\x}.\]
The first idenitity is

\noi{\bf Identity \boldmath$1_L$}: For $N\ge L$,
\[\sum_{\s\in\S_N}\prod_{1\le i<j\le N}U(\x_{\s(i)},\x_{\s(j)}){\x_{\s(2)}\,
\x_{\s(3)}^2\cd\x_{\s(N)}^{N-1}\ov(1-\x_{\s(L+1)}\cd\x_{\s(N)})\cd(1-\x_{\s(N-1)}\x_{\s(N)})\,(1-\x_{\s(N)})}\]
\be=p^{N(N-1)/2}\,{\f_L(\x)\ov\prod_i(1-\x_i)},\label{1L}\ee
where the sum is taken over all permutations $\s$ in the symmetric group $\S_N$.

For the second identity we use the notation $\xh{S}$ to denote the variables $\x_k$ with $k\ni S$, and set $\t=p/q$. We assume $\t<1$, and use the usual notation for the $\t$-binomial coefficients, 
\[\tbin{n}{k}={(1-\t^n)\cd(1-\t^{n-k+1})\ov(1-\t)\cd(1-\t^k)}.\]

\noi{\bf Identity \boldmath$2_L$}: For $0\le m\le N-L$,
\be\sum_{|S|=m}\prod_{{i\in S\atop j\ni S}}U(\x_i,\x_j)\cdot\f_L(\xh{S})=q^{m(N-m)}\tbin{N-L}{m}\,\f_L(\x),\label{2L}\ee
where the sum is over all subsets $S$ of $[1,\ld,N]$ of cardinality $m$.
\sp

For the definition of $\f_L(\x)$ we first define
\be\ph_L(z_1,\ld,z_L;\x)={\prod_{1\le j\le N}U(z_1,\x_j)\,U(z_2,\x_j)\cd U(z_L,\x_j)\ov z_1^L\,(qz_1-p)\,z_2^{L-1}\,(qz_2-p)\cd z_L\,(qz_L-p)}\,\prod_{1\le i<j\le L}{1\ov U(z_j,z_i)},\label{phidef}\ee
and then
\be\f_L(\x)=p^{L(L+1)/2-LN}\,\prod_i\x_i^L\int_{\G_\x}\cd\int_{\G_\x}\ph_L(z_1,\ld,z_L;\x)\,dz_1\cd dz_L,\label{fLdef}\ee
where $\G_\x$ consists of simple closed curves enclosing the points $\x_j$ but no other singularirites of the integrand. (Here and below all contour integrals are to get the factor $1/2\pi i$.)

By expanding the contours we can evaluate the integrals and compute the $\f_L(\x)$ explicitly. Using just $U(0,\x)=p\,\x\inv$ and $U(\t,\x)=p$
we find
\be\f_1(\x)=1-\prod_i\x_i.\label{f1}\ee
The next one is more complicated,
\[\f_2(\x)=1-e_{N-1}(\x)+{N-1\ov p}e_N(\x)-{q\ov p}\,e_1(\x)\,e_N(\x)+{q\ov p}\,e_N(\x)^2,\]
where the $e_n$ are the elementary symmetric polynomials in $N$ variables. Further $\f_L(\x)$ are even more complicated, but only for $L=1$ is it needed explicitly. From (\ref{f1}) we see that identities (\ref{1L}) and (\ref{2L}) for $L=1$ are identities (1.6) and (1.9) of \cite{tw1}, so they are true then. We shall use this fact later.
\sp

\noi{\bf Proof of Identity \boldmath$1_L$}. We perform the $z_1$-integration in (\ref{fLdef}). Using 
\[{\rm res}\,U(z,\x)|_{z=\x}=(1-\x)\,(q\x-p),\]
we find after a little computation that the integral equals
\[\sum_{k=1}^N{1-\x_k\ov\x_k^L}\,\prod_{j\ne k}U(\x_k,\x_j)\cdot\ph_{L-1}(z_2,\ld,z_L;\xh{k}),\]
where $\xh{k}$ denotes the variables other than $\x_k$.
This gives the recursion formula
\be\f_L(\x)=p^{1-N}\sum_{k=1}^N(1-\x_k)\,\prod_{j\ne k}\x_j\,U(\x_k,\x_j)\cdot\f_{L-1}(\xh{k}),\label{recursion}\ee
where we used (\ref{fLdef}) as it stands and also with $L$ replaced by $L-1$ and $N$ replaced by $N-1$. 

We prove (\ref{1L}) by induction. We know that it holds for $L=1$, so we assume that $L>1$ and that it holds for $L-1$. For fixed $k$, denote by $\S_{N,k}$ those permutations in $\S_N$ for which $\s(1)=k$. Then
\[\sum_{\s\in\S_{N,k}}\prod_{1\le i<j\le N}U(\x_{\s(i)},\x_{\s(j)}){\x_{\s(2)}\,\x_{\s(3)}^2\cd\x_{\s(N)}^{N-1}\ov(1-\x_{\s(L+1)}\cd\x_{\s(N)})\cd(1-\x_{\s(N-1)}\x_{\s(N)})\,(1-\x_{\s(N)})}\]
\[=\prod_{j\ne k}\x_j\,U(\x_k,\x_j)\sum_{\s\in\S_{N,k}}\prod_{2\le i<j\le N}U(\x_{\s(i)},\x_{\s(j)}){\x_{\s(3)}\,\x_{\s(4)}^2\cd\x_{\s(N)}^{N-2}\ov(1-\x_{\s(L+1)}\cd\x_{\s(N)})\cd(1-\x_{\s(N-1)}\x_{\s(N)})\,(1-\x_{\s(N)})}\]
By the induction hypothesis the sum over $S_{N,k}$ is equal to
\[p^{(N-1)(N-2)/2}\,{\f_{L-1}(\xh{k})\ov \prod_{i\ne k}(1-\x_i)}.\]
(Observe that $L$ changes to $L-1$ when viewed in $\S_{N,k}$.) We sum over $k$ and find that the left side of (\ref{1L}) equals  
\[{p^{(N-1)(N-2)/2}\ov\prod_i(1-\x_i)}\sum_{k=1}^N(1-\x_k)\,\prod_{j\ne k}\x_j\,U(\x_k,\x_j)\cdot\f_{L-1}(\xh{k}).\]
By (\ref{recursion}) this equals the right side of (\ref{1L}),
which completes the proof.
\sp

\noi{\bf Proof of Identity \boldmath$2_L$}. In the following $S$ will always denote a subset of $[1,\ld,N]$ with $|S|=m$. We shall use the following lemma, whose proof we give below.
\sp

\noi{\bf Lemma}. If for a set $S$ and $\l\ni S$ we define 
\be\U(\l,S)=(1-\x_\l)\(\prod_{k\in S}U(\x_k,\x_\l)-\t^{-m}\prod_{k\in S}\x_{k}\,U(\x_\l,\x_{k})\),\label{U1}\ee
then 
\[\sum_{\s\in\S_{m+1}}\U(\l,\{k_1,\ld,k_m\})=0,\]
where the sum runs over all permutations $\s$ of $(\l,k_1,\ld,k_m)$.
\sp

Assuming this, we establish (\ref{2L}) by induction on $L$. We know that it holds for $L=1$, so assume $L>1$ and that it holds for $L-1$. 

As noted, we use the notation $\xh{S}$ to denote the variables $\x_k$ with $k\ni S$. We use $\xh{S}\,\xh{\l}$ to denote the variables $\x_k$ with $k\ni S\cup\{\l\}$ (or $S\cup\l$, as we shall write it).

We use recursion formula (\ref{recursion}) to write the left side of (\ref{2L}) as
\[\sum_S\prod_{{i\in S\atop j\ni S}}U(\x_i,\x_j)\,p^{m+1-N}\sum_{\l\ni S}(1-\x_\l)\,\prod_{j\ni S\cup\l}\x_j\,U(\x_\l,\x_j)\cdot\f_{L-1}(\xh{S},\xh{\l}).\]
We write this in turn as
\be p^{m+1-N}\sum_{{S,\l\atop\l\ni S}}(1-\x_\l)
\prod_{j\ne\l}\x_j\,U(\x_\l,\x_j)\left[{\prod_{k\in S}U(\x_k,\x_\l)\ov
\prod_{k\in S}\x_k\,U(\x_\l,\x_k)}\right]
\prod_{{k\in S\atop j\ni S\cup\l}}U(\x_k,\x_j)\cdot
\f_{L-1}(\xh{S},\xh{\l}).\label{leftside}\ee

Write the expression in brackets as
\be\t^{-m}+\left[{\prod_{k\in S}U(\x_k,\x_\l)\ov
\prod_{k\in S}\x_k\,U(\x_\l,\x_k)}-\t^{-m}\right],\label{brackets}\ee
and consider the contribution of the summand $\t^{-m}$. By the induction hypothesis we have for the sum over those $S$ not containing $\l$,
\[\sum_{{S\atop\l\ni S}}\prod_{{k\in S\atop j\ni S\cup\l}}U(\x_k,\x_j)\cdot\f_{L-1}(\xh{S},\xh{\l})=q^{m(N-m-1)}\,\tbin{N-L}{m}\,\f_{L-1}(\xh{\l}).\]
If we use (\ref{recursion}) once again we see that the contribution of the $\t^{-m}$ term to  (\ref{leftside}) is exactly the right side of (\ref{2L}).  

It remains to show that if we replace the bracketed expression in (\ref{leftside}) by the one in (\ref{brackets}) the result is zero. The result of that replacement (aside from the external power of $p$) is
\[\sum_{{S,\l\atop\l\ni S}}(1-\x_\l)\Bigg[\prod_{k\in S}U(\x_k,\x_\l)-\t^{-m}\prod_{k\in S}\x_k\,U(\x_\l,\x_k)\Bigg]
\prod_{j\ni S\cup\l}\x_j\,U(\x_\l,\x_j)\prod_{{k\in S\atop j\ni S\cup\l}}U(\x_k,\x_j)\cdot\f_{L-1}(\xh{S},\xh{\l}).\]
With the notation (\ref{U1}) this may be written
\[\sum_{{S,\l\atop\l\ni S}}\U(\l,S)
\Bigg(\prod_{{k\in S\cup\l\atop j\ni S\cup\l}}\x_j\,U(\x_k,\x_j)\cdot\f_{L-1}(\xh{S},\xh{\l})\Bigg).\]
The expression in parentheses is symmetric under the permutations of $S\cup\l$. It follows from the lemma that the sum equals zero.
\sp

\noi{\bf Proof of the Lemma}. With a slightly different notation,
we show that the symmetrization of
\[(1-\xm)\(\t^m\,\prod_{i=1}^{m} U(\x_i,\xm)-\prod_{i=1}^{m}\x_i\,U(\xm,\x_i)\)\]
over $S_{m+1}$ equals zero. The symmetrization is equal to $(m+1)\inv$ times
\be\sum_{j=1}^{m+1}(1-\x_j)\(\t^m\,\prod_{i\ne j}U(\x_i,\x_j)-\prod_{i\ne j}\x_i\,U(\x_j,\x_i)\).\label{id}\ee
To show that this equals zero, we consider the integral
\[\int{1\ov p-qz}\;\Big[\t^m\prod_{i=1}^{m+1} U(\x_i,z)+z\inv\prod_{i=1}^{m+1}\x_i\,U(z,\x_i)\Big]\,dz\]
taken over a large contour. The integrand equals $=-p^m\prod\x_i\, z\inv+O(z^{-2})$ as $z\to\iy$, so the integral equals
\be-p^m\prod\x_i.\label{integral}\ee 
This equals the sum of the residues of the integrand. Using
\be{\rm res}\ U(\x,z)|_{z=\x}=(1-\x)(p-q\x),\ \ \ {\rm res}\ U(z,\x)|_{z=\x}=-(1-\x)(p-q\x),\label{residues}\ee
\[U(0,\x)=p\,\x\inv,\ \ \ U(\x,\t)=q,\ \ \ U(\t,\x)=p.\]
we find that the sum of the residues at the $\x_j$ equals (\ref{id}), the residue at $z=0$ equals $p^m$, and the residue at $z=\t$ equals $-p^m-p^m\prod\x_i$. That these residues add up to (\ref{integral}) is equivalent to (\ref{id}) being equal to zero.
\sp

\noi{\bf Remark}. We stated earlier that $\f_L(\x)$ is a symmetric polynomial. Symmetry is clear from the definition, but that it is a polynomial is not so clear. Here is an inductive argument, using (\ref{recursion}). 

It is a rational function of each $\x_i$ whose poles are among the $\x_j$ with $j\ne i$. We show that the residue at $\x_i=\x_j$ is zero. 
The $k$-summand in (\ref{recursion}) with $k\ne i,\,j$ is analytic at $\x_i=\x_j$, since there is no denominator $\x_i-\x_j$ or $\x_j-\x_i$. The $k=i$ summand has a pole coming from the factor $U(\x_i,\x_j)$ and no other factor, while the $k=j$ summand has a pole coming from the factor $U(\x_j,\x_i)$ and no other factor. Using (\ref{residues}) and the symmetry of $\f_{L-1}$ we see that the two residues are negatives of each other, and so their sum is zero.

\bc{\bf III. Block probabilities}\ec

For $N$-particle ASEP we denote by $\P_Y(X,t)$ the probability that a systenm initially in configuration $Y=\{y_1,\ld,y_N\}$ is in configuration $X=\{x_1,\ld,x_N\}$ at time~$t$. Other notation is
\be\ep(\x)=p\,\x\inv+q\,\x-1,\ \ \ A_\s(\x_1,\ld,\x_N)=\prod_{i<j}{U(\xs{i},\xs{j})\ov U(\x_i,\x_j)}.\label{A}\ee
(The expression for $A_\s$ here is easily seen to agree with formula (3.2) of \cite{tw1}.) Theorem~2.1 of \cite{tw1} is that for $p>0$,
\be\P_Y(X,t)=\sum_{\s\in\S_N}\int_{\C_r}\cd\int_{\C_r}A_\s\,\prod_i\(\xs{i}^{x_i-\ys{i}-1}\,e^{\ep(\x_i)t}\)\,d\x_1\cd\x_N,\label{PXY}\ee
where $\C_r$ is a circle with center zero and radius $r$ so small that all poles of the $A_\s$ lie outside $\C_r$.

For blocks of length $L$ we define $\P_{L,\,Y}(x,m,t)$ to be the probability that at time~$t$ the $m$th particle from the left is the beginning of a block starting at $x$. Precisely, that 
\[x_m(t)=x,\ x_{m+1}(t)=x+1,\ld,x_{m+L-1}(t)=x+L-1.\]
To state our basic result, we define
\[I_L(x,Y,\x)=\prod_{i<j}{1\ov U(\x_i,\x_j)}\prod_i{1\ov1-\x_i}\,\f_L(\x)\prod_i\(\x_i^{x-y_i-1}\,e^{\ep(\x_i)t}),\]
where all indices lie in $[1,\ld,N]$. For a set $S\subset[1,\ld,N]$ we define $I_L(x,Y_S,\x_S)$ analogously, where the indices lie in $S$. We denote by $S^c$ the complement of $S$ in [1,\ld,N] and, finally, we confuse things by defining $\s(S^c)$ to be the sum of the elements of $S^c$. The result is
\sp

\noi{\bf Theorem 1}. For $p>0$,
\[\P_{L,Y}(x,m,t)=p^{(N-m+1)(N-m)/2}\,q^{(m-1)(N-m/2)}\sum_{|S^c|<m}(-1)^{m-1-|S^c|}\,\tbin{|S|-L}{m-1-|S^c|}\]
\[\times\, {q^{\s(S^c)-N|S^c|}\ov p^{\s(S^c)-|S^c|(|S^c|+1)/2}}\,\int_{\C_r}\cd\int_{\C_r}I_L(x,Y_S,\x_S)\,d^{|S|}\x.\]

We shall only give the details for the cases $m=1$ and $m=2$, from which one can see how identities $1_L$ and $2_L$ replace those for $L=1$ in \cite{tw1}. Because the $\f_L(\x)$ are polynomials there are no extra complications following the derivation in \cite{tw1} since no new poles are introduced. There are, though, other properties of $\f_L(\x)$ needed for the applications of Lemmas 3.1 and 5.1 of \cite{tw1}.  We derive these in Appendix A.

For $m=1$ the formula is
\be\P_{L,\,Y}(x,1,t)=p^{N(N-1)/2}\int_{\C_r}\cd\int_{\C_r}I_L(x,Y,\x)\,d^N\x.\label{m=1}\ee
To see this we write the configuration that starts at $x$ as
\[x,\,x+1,\ld,x+L-1,\,x+L-1+v_1,\ld,x+L-1+v_1+\cd+ v_{N-L},\]
and then sum the right side of (\ref{PXY}) over all $v_i>0$. After summing, the integrand in (\ref{PXY}) becomes
\[A_\s\,{\x_{\s(2)}\,
\x_{\s(3)}^2\cd\x_{\s(N)}^{N-1}\ov(1-\x_{\s(L+1)}\cd\x_{\s(N)})\cd(1-\x_{\s(N-1)}\x_{\s(N)})\,(1-\x_{\s(N)})}\,\prod_i\(\x_i^{x-y_i-1}\,e^{\ep(\x_i)t}),\]
and using the definition of $A_\s$ in (\ref{A}) we see that (\ref{m=1}) follows from Identity $1_L$.
\sp

For $m=2$ we denote by $\wh{Y_k}$ the set $Y\backslash\{k\}$, and similarly for $\xh{k}$ as before. The formula is
\pagebreak

\[\P_{L,\,Y}(x,2,t)=-q^{N-1}\,\tbin{N-L}{1}\,p^{(N-1)(N-2)/2}\int_{\C_r}\cd\int_{\C_r}I_L(x,Y,\x)\,d^N\x\]
\be+\,p^{(N-1)(N-2)/2}\sum_{k=1}^N\({q\ov p}\)^{k-1}\int_{\C_r}\cd\int_{\C_r}I_L(x,\wh{Y_k},\xh{k})\,d^{N-1}\x.\label{m=2}\ee
For this we write the configuration as
\[x-v_1,\,x,\,x+1,\ld,x+L-1,\,x+L-1+v_2,\ld,x+L-1+v_1+\cd+ v_{N-L},\]
with $v_i>0$. Summing the integrand over $v_2,\ld,v_{N-L}$ gives for the factors involving~$\s$
\be A_\s\;\x_{\s(1)}^{-v_1}\;{\x_{\s(3)}\,
\x_{\s(4)}^2\cd\x_{\s(N)}^{N-2}\ov(1-\xs{L+2}\xs{L+3}\cd\xs{N})\cd(1-\xs{N-1}\xs{N})\,(1-\xs{N})}.\label{Aprod}\ee
To do the sum over $v_1$ we expand the $\xs{1}$-contour to $\C_R$ with large $R$. As in \cite{tw1} no poles are passed in the contour deformation. The factor $\xs{1}^{-v_1}$ becomes $1/(\xs{1}-1)$ after the summation, whereupon we deform the $\xs{1}$-contour back to $\C_r$. In this deformation,  we pass the pole at $\xs{1}=1$, the value of $A_\s$ there being
\[\({q\ov p}\)^{k-1}\;{\ds{\prod_{1<i<j}U(\xs{i},\xs{j})}\ov \ds{\prod_{{i<j\atop i,j\ne k}}U(\x_i,\x_j)}}\]
when $\s(1)=k$. Now we think of $\s$ as a function from $[2,\ld,N]$ to $[1,\ld,N]\bs\{k\}$. We multiply by the quotient in (\ref{Aprod}), sum over those $\s$ with $\s(1)=k$, and use identity~$1_L$ (with obvious modification) to obtain
\[\({q\ov p}\)^{k-1}\;p^{(N-1)(N-2)}\,\prod_{{i<j\atop i,j\ne k}}{1\ov U(\x_i,\x_j)}\,\prod_{j\ne k}{1\ov 1-\x_j}\;\f_{L-1}(\xh{k}).\]
The exterior factor is now $\prod_{i\ne k}\Big(\x_i^{x-y_i-1}\,e^{\ep(\x_i)t}\Big),$
and from these, summing on $k$, we obtain the sum in (\ref{m=2}).

This is the contribution from the residues when we deform the $\xs{1}$-contour. Now we consider the integral we have after the contour deformations and summation over~$v_1$. Again we consider at first only the summands in which $\s(1)=k$. The factor $A_\s/(\xs{1}-1)$ becomes
\[{1\ov\x_k-1}{\prod_{j\ne k}U(\x_k,\x_j)\ov\prod_{i<j}U(\x_i,\x_j)}\,
\prod_{1<i<j}U(\xs{i},\xs{j}).\]
Then as before we multiply by the quotient in (\ref{Aprod}), sum over those $\s$ with $\s(1)=k$, and use identity $1_L$ to obtain
\[-p^{(N-1)(N-2)}\,\prod_{j=1}^N{1\ov1-\x_j}\;{\prod_{j\ne k}U(\x_k,\x_j)\ov\prod_{i<j}U(\x_i,\x_j)}\,\f_{L-1}(\xh{k}).\]
Finally, to sum over $k$ we apply identity $2_L$ and obtain
\[-q^{N-1}\,\tbin{N-k}{1}\,p^{(N-1)(N-2)}\,\prod_{j=1}^N{1\ov1-\x_j}\,\prod_{i<j}{1\ov U(\x_i,\x_j)}\;\f_L(\x).\]
Multiplying by the factor $\prod_i\Big(\x_i^{x-y_i-1}e^{\ep(\x_i)t}\Big)$ gives the first term in (\ref{m=2}).

The next result is a formula for the same probability but with integrations over large contours.  

\noi{\bf Theorem 2}. For $q>0$,
\[\P_{L,\,Y}(x,m,t)=(-1)^{m+1}\,p^{m(m-1)/2}
\sum_{|S|\ge m+L-1} q^{(m-1)(|S|-m/2)}\,\tbin{|S|-L}{m-1}\]
\be\times{p^{\s(S)-m|S|}\ov q^{\s(S)-|S|(|S|+1)/2}}\int_{\C_R}\cd\int_{\C_R}I_L(x,Y_S,\x_S)\,d^{|S|}\x,\label{Th2}\ee
where $R$ is so large that the poles of the integrand lie inside $\C_R$.

In \cite{tw1} we used a duality between our ASEP and one with $p$ and $q$ interchanged to derive Theorem 2 quickly from Theorem 1 in the case $L=1$. For the argument to extend to general $L$ we would need the following:

If we make the replacments $\t\to 1/\t,\ \x_i\to\x_i\inv$ in $\f_L(\x)$ the result is equal to
\[(-1)^L\,{\t^{L(L-1)/2}\ov\prod\x_i^L}\,\f_L(\x).\]
Although this has been verified in many case we have not (yet) found a proof. Instead we use Lemma 3.1 of \cite{tw1}, which expresses an integral of the type we have over small contours in terms of integrals over large contours. This more elaborate argument is presented in Appendix B. 

As in \cite{tw1}, Theorem 2 extends to infinite systems unbounded on the right. The sum is then taken over finite subsets of $\Z^+$.
\pagebreak

\bc{\bf IV. Step initial condition}\ec

As in \cite{tw2}, with step initial condition ($Y=\Z^+$) there is an expression for the probability $\P_{L,\,\Z^+}(x,m,t)$ as an integral involving a Fredholm determinant. Before stating the result we derive an alternative expression for the integral in (\ref{fLdef}), namely
\be\int_{\G_\x}\cd\int_{\G_\x}\ph_L(z_1,\ld,z_L;\x)\,dz_1\cd dz_L=
(-1)^L\,\int_{\G_{0,\t}}\cd\int_{\G_{0,\t}}\ph_L(z_1,\ld,z_L;\x)\,dz_L\cd dz_1.\label{2ints}\ee
Informally, $\G_{0,\t}$ is a contour consisting of tiny circles around the points $z=0$ and $z=\t$, with the circles for each $z_{i}$ lying well outside the circles for $z_{i+1}$. Precisely, the iterated integral on the right is interpreted as follows: First take the sum of the residues at $z_L=0$ and $z_L=\t$. In the resulting integrand take the sum of the residues at $z_{L-1}=0$ and $z_{L-1}=\t$. And so on. 

Here is the argument for (\ref{2ints}). First evaluate the $z_L$-integral on the left by expanding the contour. There are contributions from minus the residues at 0 and $\t$, none from infinity, and one from minus the residue at each pole at $z_L=p/(1-qz_i)$. The factors involving $z_i$ when we compute the latter residue combine as a constant times
\[{(1-z_i)\ov z_i^{L-i+1}(1-pz_i)}\prod_j{z_i-\x_j\ov \x_j-qz_i\x_j-p}.\]
This no longer has the singularities at the $z_i=\x_j$. Therefore the  
$z_i$-integral over $\G_\x$ equals zero, which means that the residue at  $z_L=p/(1-qz_i)$ integrates to zero. Thus the integral with respect to $z_L$ over $\G_\x$ equals minus the sum of the residues at $z_L=0$ and $z_L=\t$. Then we find similarly that the resulting integral with respect to $z_{L-1}$ over $\G_\x$ equals minus the sum of the residues at $0$ and $\t$. Continuing this way we replace all integrals with respect to the $z_i$ over 
$\G_\x$ by minus the sum of the residues at $0$ and~$\t$. (And we have to do it in the order $z_L,\, z_{L-1},\, \ld, z_1$.)

With (\ref{2ints}) established, we introduce the notation $K_{L,x}(z)$ for the integral operator acting on functions on $\C_R$ with kernel
\[K_{L,\,x}(\x,\x';\,z)=K_x(\x,\x')\,\prod_{j=1}^LU(z_j,\x),\]
where
\[K_x(\x,\x')={\x^x\,e^{\ep(\x)t}\ov p+q\x\x'-\x}.\]
In the statement below the integral over the $z_i$ is interpreted as in the right side of (\ref{2ints}), and $(\la;\t)_m$ is the Pochhammer symbol $\prod_{j=0}^{m-1}(1-\la\,\t^j)$.
\pagebreak

\noi{\bf Theorem 3}. For $p,\,q>0$,
\[\P_{L,\,\Z^+}(x,m,t)=(-1)^{L-1}\,p^{L(L+1)/2}\,\t^{-(m-1)(L-1)}\]
\[\times\int_{\G_{0,\t}}\cd\int_{\G_{0,\t}}{1\ov z_1^L\,(qz_1-p)\,z_2^{L-1}\,(qz_2-p)\cd z_L\,(qz_L-p)}\,\prod_{i<j}{1\ov U(z_j,z_i)}\]
\[\times\,\left[\int{\det(I-p^{-L}q\,\la\,K_{L,\,x+L-1}(z))\ov (\la;\t)_m}\;{d\la\ov\la^L}\right]\,dz_L\cd dz_1\,.\]
The $\la$-integration is over a contour enclosing the singularities of the integrand at $\t^{-j}$ for $j=0,\ld,m-1$.

\noi{\bf Remark}. For the case $L=1$, evaluating the $z_1$ integral by computing residues at 0 and $\t$ one obtains
\[\P_{1,\Z^+}(x,m,t)=\int{\det(1-q\la K_x)-\det(1-q\la K_{x-1})\ov(\la;\t)_m}\,{d\la\ov\la}.\]
This implies
\[\mathbb P_{\Z^+}(x_m(t)\le x)=\int{\det(1-q\la K_x)\ov(\la;\t)_m}\,{d\la\ov\la},\]
which is equation (1) of \cite{tw2}.

For the proof of Theorem 3, we first simplify (\ref{Th2}) when $Y=\Z^+$ as in \cite{tw1} by first summing over all $S$ with $|S|$ equal to a fixed $k$. We define
\[ J_{L,k}(x,t,\x)=\prod_{i\ne j}{1\ov U(\x_i,\x_j)}\,{\f_L(\x)\ov\prod_i(1-\x_i)(q\x_i-p)}\,\prod_i \x_i^{x-1}\,e^{\ep(\x_i)\,t},\]
where the indices run over $[1,\ld,k]$. The result is
\[\P_{L,\,\Z^+}(x,m,t)=(-1)^{m+1}\sum_{k\ge m+L-1}{1\ov k!}\,\tbin{k-L}{m-1}\]
\be\times\,p^{(k-m)(k-m+1)/2}\,q^{km+(k-m)(k+m-1)/2}\,
\int_{\C_R}\cd\int_{\C_R}J_{L,k}(x,t,\x)\,d\x_1\cd\x_k.\label{stepprob}\ee
The derivation follows the same steps as in \cite{tw1}.

Using (\ref{phidef})--(\ref{fLdef}), which give the definition of $\f_L(\x)$, and (\ref{2ints}), we obtain
\pagebreak

\[J_{L,k}(x,t,\x)=(-1)^L\,p^{-kL+L(L+1)/2}\int_{\G_\x}\cd\int_{\G_\x}{1\ov z_1^L\,(qz_1-p)\,z_2^{L-1}\,(qz_2-p)\cd z_L\,(qz_L-p)}\,\prod_{i<j}{1\ov U(z_j,z_i)}\]
\be\times\left[\prod_{i\ne j}{1\ov U(\x_i,\x_j)}\,\prod_i{U(z_1,\x_i)\cd U(z_L,\x_i)\ov (1-\x_i)(q\x_i-p)}\,\prod_i\x_i^{x+L-1}\,e^{\ep(\x_i)t}\right]\,dz_L\cd dz_1.\label{Jkrep}\ee

We saw in \cite{tw2} that 
\[\det(K_x(\x_i,\x_j))_{i,j\le k}=(-1)^k\,(pq)^{k(k-1)/2}\,
\prod_{i\ne j}{1\ov U(\x_i,\x_j)}\,\prod_i{1\ov(1-\x_i)(q\x_i-p)}\,\prod_i\x_i^x\,e^{\ep(\x_i)t}.\]
It follows that the expression in brackets in (\ref{Jkrep}) equals
\[(-1)^k\,(pq)^{-k(k-1)/2}\,\det(K_{L,\,x+L-1}(\x_i,\x_j;\,z))_{i,j\le k}.\]
Therefore (\ref{Jkrep}) becomes
\[J_{L,k}(x,t,\x)=(-1)^{k+L}\,p^{-kL+L(L+1)/2}(pq)^{-k(k-1)/2}\,\int_{\G_{0,\t}}\cd\int_{\G_{0,\t}}{1\ov z_1^L\,(qz_1-p)\,z_2^{L-1}\,(qz_2-p)\cd z_L\,(qz_L-p)}\]
\be\times\,\prod_{i<j}{1\ov U(z_j,z_i)}\ \det(K_{L,\,x+L-1}(\x_i,\x_j;\,z))_{i,j\le k}\ dz_L\cd dz_1.\label{J1}\ee

Next, we obeserve that
\[{(-1)^k\ov k!}\,\int_{\C_R}\cd\int_{\C_R}\det(K_{L,\,x+L-1}(\x_i,\x_j;\,z))_{i,j\le k}\,d\x_1\cd d\x_k=\int{\det(I-\la\,
K_{L,\,x+L-1}(z))\ov \la^{k+1}}\,d\la\]
\[=p^{kL}\,\int{\det(I-p^{-L}\la\,K_{L,\,x+L-1}(z))\ov \la^{k+1}}\,d\la\]
the $k$-th coefficient in the Fredholm expansion of $\det(I-\la\,
K_{L,\,x+L-1}(z))$.

The final step is to sum over $k$, interchanging the sum with the integrals. The sum of the terms involving $k$ equals
\[\sum_{k\ge L+m-1}^\iy p^{(k-m)(k-m+1)/2}\,q^{km+(k-m)(k+m-1)/2}\,(pq)^{-k(k-1)/2}\,\la^{-k}\,\tbin{k-L}{k-L-m+1}\]
\be=p^{L+m-1}\,\t^{-Lm-m(m-1)/2}\,\la^{-L-m+1}\sum_{k\ge0}p^k\,\t^{-mk}\,\la^{-k}\tbin{k+m-1}{k}.\label{preceding}\ee
By the $\t$-binomial theorem we have for $|z|<1$,
\[\sum_{k\ge0}z^k\tbin{k+m-1}{k}=\prod_{j=0}^{m-1}(1-z\,\t^j)\inv,\]
from which it follows that for large enough $\la$ (\ref{preceding}) is equal~to
\[(-1)^m\,p^{L-1}\t^{-(L-1)m}\,\la^{-L+1}\,{1\ov(\la\,q\inv;\,\t)_m}.\]
After making the substitution $\la\to q\la$ in the $\la$-integral and referring back to (\ref{stepprob}) and (\ref{J1}), we see that Theorem 3 follows.

\bc{\bf Appendix A}\ec

For our applications of Lemmas 3.1 and 5.1 of \cite{tw1} one has to know two things. The first is that 
\[\f_L(\x)|_{\x_k=1}=\f_{L-1}(\xh{k}).\]
This follow from (\ref{fLdef}) and the fact $U(z,1)=p$.

The second property, for the application of Lemma 5.1, is that for $i\ne k$ we have $\f_L(\x)=O(1)$ as $\x_k\to\iy$ when $\x_i=p/(1-q\x_k)$ and the other $\x_j$ are bounded. (For the application of Lemma 3.1 one intechanges $\x_i$ and $\x_k$.) This is clearly true for $L=1$ and we use induction to prove it for $L>1$. 
So as not to confuse indices we write (\ref{recursion}) as
\[\f_L(\x)=p^{1-N}\sum_{\l=1}^L(1-\x_\l)\,\prod_{j\ne \l}\x_j\,U(\x_\l,\x_j)\cdot\f_{L-1}(\xh{\l}).\]

Because $\f_L(\x)$ is a polynomial we may assume that the $\x_\l$ with $\l\ne i,k$ lie on circles centered at zero with different radii. The reason is that the maximum modulus theorem, applied one variable at a time, would extend the bound to inside these circles, which are arbitrarily large. With this assumption all $U(\x_\l,\x_j)=O(1)$, no matter what the indices.

The $\l=i$ summand equals zero because the product over $j$ contains the factor
\[U(\x_i,\x_k)=U(p/(1-q\x_k),\,\x_k)=0.\]
For $\l\ne i$ the factors $\f_{L-1}(\xh{\l})$ are $O(1)$ by induction hypothesis since $\x_i$ is one of the remaining variables. Then for $\l=k$ there is the factor $1-\x_k$ but the product over $j$ contains the factor $\x_i$ but not $\x_k$ and so is $O(\x_k\inv)$. Finally, when $\l\ne i,k$ the product over $j$ contains both factors $\x_k$ and $\x_i$, and their product is $O(1)$. So all summands are $O(1)$.

\bc{\bf Appendix B}\ec

Here we derive Theorem 2 from Theorem 1 using Lemma 3.1 of \cite{tw1}. The right side of Theorem 1 is a sum over $S\subset[1,\ld,N]$ of integrals with variables $\x_S$ with coefficients
\be (-1)^{m+N+1}\,\t^{N(N+1)/2+m(m-1)/2-mN}\label{coeff1}\ee
(which are independent of $S$) times
\be (-1)^{|S|}\,\tbin{|S|-L}{N-m-L+1}\label{coeff2}\ee
times
\be\t^{-N|S|}\,\t^{\s(S)}\,p^{|S|(|S|-1)/2}.\label{coeff3}\ee

We apply Lemma 3.1 of \cite{tw1}, with $[1,\ld,N]$ replaced by $S$, to the integrals of Theorem 1. (In Appendix A we showed that the hypothesis of Lemma 3.1 holds here.) The statement of that lemma uses the notation, for sets $U$ and $V$,
\[\s(U,V)=\#\{(i,j):i\ge j,\ i\in U,\ j\in V\}.\]
If $V=[1,\ld,N]$ this equals $\s(S)$. We shall use the fact that $\s(U,V)$ is linear in both $U$ and $V$.

Applying the lemma we get integrals over sets $T\subset S$ of integrals with integrands $I_K(x,Y_T,\x_T)$, and with coefficients
\[{p^{|S\bs T|-\s(S\bs T,\,S)}\ov q^{\s(T,\,S)-|T|(|T+1)/2)}}.\]
The product of this with (\ref{coeff3}) is equal to
\be{p^{-|T|}q^{|T|(|T|+1)/2}}\,\t^{\s(T)}\,\t^{|S|(|S|+1)/2-N|S|}\,\t^{\s(S\bs T)-\s(S\bs T,\,S)}.\label{coeff4}\ee
Eventually we are going to get a sum over $T$ (which will replace $S$ in the statement of Theorem 2), so we fix $T$ and then sum over all $S$ satisfying
\[[1,\ld,N]\supset S\supset T.\]
The first three factors in (\ref{coeff4}) depend on $T$ only, the next factor depends only on $|S|$ and the last depends on $S$. First we fix $k$ and sum over all $S$ with $|S|=k$. So at first we need only sum the last factor (leaving the sum over $k$ for later). The sum is
\[\sum_{{[1,\ld,N]\supset S\supset T\atop|S|=k}}\t^{\s(S\bs T)-\s(S\bs T,\,S)}.\]
With $T^c=[1,\ld,N]\bs T$ we have 
\[\s(S\bs T)-\s(S\bs T,\,S)=\s(S\bs T,\,T)+\s(S\bs T,\,T^c)-
\s(S\bs T,\,S\bs T)-\s(S\bs T,\,T)\]
\[=\s(S\bs T,\,T^c)-\s(S\bs T,\,S\bs T).\]
Replacing $S\bs T$ by $S$, we may rewrite our sum as
\[\sum_{{S\subset T^c\atop |S|=k-|T|}}\t^{\s(S,\,T^c)-\s(S,\,S)}.\]
Using the order-preserving mapping $T^c\to [1,\ld,N-|T|\,]$ we see that the sum equals
\[\sum_{{S\subset[1,\ld,N-|T\,|]\atop|S|=k-|T|}}\t^{\s(S)-\s(S,\,S)}.\]
This equals \cite{KC} 
\[\tbin{N-|T|}{k-|T|}.\]

If we multiply this by (\ref{coeff2}) and the factor $\t^{|S|(|S|+1)/2-N|S|}$ from (\ref{coeff4}) and sum over $k$ we get
\be\sum_{k=|T|}^N(-1)^k\,\t^{k(k+1)/2-Nk}\,\tbin{k-L}{N-m-L+1}\,\tbin{N-|T|}{k-|T|}.\label{ksum}\ee

Taking into account (\ref{coeff1}) and the powers of $p$ and $q$ in (\ref{coeff4}) we find that this would agree with Theorem~2 if the sum were equal to
\be(-1)^N\,\t^{-N(N+1)/2+mN-(m-1)|T|}\,\tbin{|T|-L}{m-1}.\label{sum?}\ee

To show that this is so, denote the sum by $F(m)$. First, we have
\be F(1)=(-1)^N\,\t^{-N(N-1)/2}.\label{F1}\ee
(Observe that when $m=1$ the only nonzero summand in (\ref{ksum}) occurs when $k=N$.) Second, the algorithm \texttt{qZeil} \cite{Z} produces the recursion formula
\[F(m)={\t^{N-L-m+2}\,(1-\t^{L+m-|T|-2})\ov 1-\t^{m-1}}\,F(m-1).\]
Taking the product and using (\ref{F1}) give
\[F(m)=(-1)^{N+m-1}\,\t^{-N(N+1)/2}\,\prod_{j=2}^m{\t^{N-L-j+2}\,
(1-\t^{L+j-|T|-2})\ov 1-\t^{j-1}}\]
\[=(-1)^N\,\t^{-N(N+1)/2+mN-(m-1)|T|}\,\prod_{j=2}^m\,{1-\t^{|T|-L-j+2}\ov 1-\t^{j-1}},\]
which is (\ref{sum?}).

\bc{\bf Acknowledgment}\ec

This work was supported by the National Science Foundation through grants DMS--1207995 (first author) and DMS--1400248 (second author).

\end{document}